\documentclass{kluwer}    % Specifies the document style.

\newdisplay{guess}{Conjecture}

\begin{document}                                                                                   
\begin{article}
\begin{opening}         

\title{Do the Unidentified EGRET Sources Trace Annihilating Dark Matter 
in the Local Group?}

\author{J. \surname{Flix}$^{1}$ \thanks{Contact e-mail address: \email{jflix@ifae.es}}}
\author{J.E. \surname{Taylor}$^{2}$}
\author{M. \surname{Mart{\'i}nez$^{1}$}}
\author{F. \surname{Prada$^{3}$}}
\author{J. \surname{Silk$^{2}$}}
\author{J. \surname{Cortina$^{1}$}}

\runningauthor{J. Flix et al} \runningtitle{Do the Unidentified EGRET Sources
Trace Dark Matter?}

\institute{$^{1}$ Institut de F{\'i}sica d'Altes Energies, Universitat
  Aut{\`o}noma de Barcelona, E-08193 Barcelona, Spain \\ $^{2}$
  Astrophysics, University of Oxford, Denys Wilkinson Building, Keble
  Road, Oxford, OX1 3RH, UK \\ $^{3}$ Instituto de Astrof{\'i}sica de
  Andaluc{\'i}a (CSIC), E-18008 Granada, Spain}

\begin{abstract}
In a cold dark matter (CDM) framework of structure formation, the dark matter
haloes around galaxies assemble through successive mergers with smaller haloes. 
This merging process is not completely efficient, and hundreds
of surviving halo cores, or {\it subhaloes}, are expected to remain in
orbit within the halo of a galaxy like the Milky Way. While the dozen
visible satellites of the Milky Way may trace some of these subhaloes, 
the majority are currently undetected. A large number of high-velocity 
clouds (HVCs) of neutral hydrogen {\it are} observed around the Milky Way, 
and it is plausible that some of the HVCs may trace subhaloes undetected 
in the optical. Confirming the existence of concentrations of dark matter 
associated with even a few of the HVCs would represent a dramatic step 
forward in our attempts to understand the nature of dark matter.
Supersymmetric (SUSY) extensions of the Standard Model of particle physics 
currently suggest neutralinos as a natural well-motivated candidate for the 
non-baryonic dark matter of the universe. If this is indeed the case, then 
it may be possible to detect dark matter indirectly as it annihilates into
neutrinos, photons or positrons. In particular, the centres of subhaloes 
might show up as point sources in gamma-ray observations. In this work we 
consider the possibility that some of the unidentified EGRET $\gamma$-ray 
sources trace annihilating neutralino dark matter in the dark substructure 
of the Local Group. We compare the observed positions and fluxes of both
the unidentified EGRET sources and the HVCs with the positions and fluxes
predicted by a model of halo substructure, to determine to what extent
any of these three populations could be associated.
\end{abstract}

\end{opening}

\section{Introduction} 

Determining the nature of the unidentified EGRET sources \cite{HartmanEtAl} 
is an important unsolved problem in modern high-energy Astrophysics. These 
sources constitute a major part of all the discrete sources sample detected 
by EGRET. Despite continuing efforts, most of them have not yet been 
associated with sources at other wavelengths. Identification of EGRET 
sources is particularly difficult because of the large uncertainties in 
source positions and the lack of other observations at similar wavelengths.

While a few of the unidentified EGRET sources may simply be artifacts
of the Galactic diffuse emission, it is normally thought that most are
genuine detections, and that some fraction may even represent a
completely new population of $\gamma$-ray emitters. As the sources 
are located predominantly at low to mid-Galactic latitudes, it seems
likely they are mainly Galactic in origin, rather than being extragalactic 
objects like quasars (although this may account for some faction of them 
-- e.g.\ \cite{SowardsEtAl}). Thus most of the effort to identify the EGRET 
sources has focused on possible Galactic counterparts such as young pulsars 
\cite{HobbsetAl}, supernova remnants \cite{Torres}, isolated neutron 
stars \cite{Goldoni}, molecular clouds \cite{Benaglia}, objects in Gould's 
Belt \cite{Gehrels}, microquasars \cite{Paredes}, or young stellar objects 
like Wolf-Rayet stars or OB associations \cite{RomeroetAl}.

On extragalactic scales, observational cosmology has recently converged 
on a `concordance' model, which appears to satisfy all or most of the 
observational constraints simultaneously. In this `$\Lambda CDM$' cosmology, 
the present-day energy density of the universe is dominated by `dark energy', 
possibly in the form of a cosmological constant $\Lambda$, while the
remaining matter content is constrained by nucleosynthesis and the microwave
background to be predominantly non-baryonic cold dark matter (CDM) \cite{WMAP}.
The nature of this dark matter is still unknown, and its identification is
currently a top priority for theoretical and experimental Astrophysics. 

Weakly Interacting Massive Particles (WIMPs), non-relativistic relics
from the early universe, have been proposed as one good candidate for
non-baryonic dark matter. Although none of the particles in the
Standard Model satisfy the requirements of being weakly interacting, 
massive and stable enough to be suitable relic particle, supersymmetric 
(SUSY) extensions of the Standard Model do provide a candidate that
satisfies these requirements: the lightest SUSY particle (LSP), which
ends up being one of the neutralinos ($\chi$) in most of the SUSY-breaking
scenarios. The neutralino is a majorana particle and can annihilate
with itself in pairs, producing high-energy neutrinos, positrons,
antiprotons and $\gamma$-rays \cite{BergstromEtAl}. Thus in principle 
it can be detected through high-energy observations.

In this paper we will discuss about the possibility that $\gamma$-ray
emission from $\chi$ self-annihilation, occurring in dark matter
subhaloes in the halo of our Galaxy, may be responsible for some of
the unidentified EGRET sources. We consider the spatial distribution
and fluxes of these sources, and compare them with the predictions
of a model of halo substructure. We will also compare the source
positions with the high-velocity clouds (HVCs), a population of
neutral hydrogen clouds that may also trace dark matter concentrations 
in the galactic halo.

\section{Dark Matter Substructure in Galaxy Haloes}

In CDM cosmologies, galaxies form within dense concentrations of 
dark matter, called `haloes' by analogy with the stellar halo of the
Milky Way. Dark matter haloes are assembled through a process of 
accretion and the hierarchical merging of smaller subunits (cf.\ 
\cite{TaylorEtAl} and references therein). High-resolution numerical
simulations of CDM structure formation have shown that this merging 
process is not completely efficient; they predict that hundreds of 
dense halo cores, or {\it subhaloes}, should remain in orbit around 
present-day galaxies (e.g.\ \cite{Klypin}). While some of these 
subhaloes may correspond to the few dozen satellites which surround 
the Milky Way Galaxy and our neighbour M31 \cite{Mateo}, the 
expected number of satellites exceeds the number observed by 
approximately an order of magnitude \cite{Klypin} \cite{Moore2}. 

The current census of dwarf satellites is probably still incomplete. 
The local dwarf galaxies are often faint, low-brightness objects, and 
thus very difficult to detect, as evidenced by several discoveries
made only recently (e.g.\ the Canis Major dwarf, \cite{MartinEtAl}). 
Nonetheless, it seems unlikely that most of the subhaloes predicted 
to orbit the Milky Way will ever be detectable by stellar emission alone.

\subsection{Subhaloes as $\gamma$-ray Sources}

The specific properties of the dark matter particle may provide an
alternate  means of detecting local substructure. Neutralino
annihilation can  generate continuum $\gamma$-ray emission,
via the process  $\chi\chi\rightarrow q \bar{q}$. The subsequent decay
of $\pi^{0}$-mesons created in the resulting quark jets produces
a continuum of $\gamma$-rays.  Given a neutralino density profile
$\rho_{\rm dm}(r)$ (i.e. the dark matter density  profile within a
subhalo), the expected annihilation $\gamma$-ray flux above  an energy
$E_{0}$ arriving to Earth is given by:

\begin{eqnarray}
\Phi_{\gamma} (E>E_{0}) = \frac{1}{4\pi} \frac{N_{\gamma}\left<\sigma v\right>_{\rm ann}}{2
 m_{\chi}^{2}} \cdot J(\Psi)
\label{Eq1}
\end{eqnarray} 

\begin{eqnarray}
J(\Psi) = \int_{\rm l.o.s} \rho^{2}_{\rm dm}(r) dl
\label{Eq2}
\end{eqnarray}

where $N_{\gamma}$ is the number of continuum gammas emitted per
neutralino annihilation of $m_{\chi}$ mass above an energy $E_{0}$,
and $\left<\sigma v\right>_{\rm ann}$ the averaged product of
annihilation cross-section and relative velocities. The first part of
Equation~\ref{Eq1} can be evaluated for any given SUSY scenario,
although the result varies by several orders of magnitude over the
span of SUSY parameter space  \cite{PradaEtAl}.

All the information about the dark matter distribution is contained
in the $J(\Psi)$ term. Cosmological simulations indicate that dark matter
haloes are approximately spherical, with a cuspy central density profile
that scales as $\rho_{\rm dm}(r) \propto r^{-\alpha}$. The central slope 
of the density profile is thought to lie somewhere between $\alpha$=1 
\cite{NavarroEtAl} (the `NFW profile') and $\alpha$=1.5 \cite{MooreEtAl}, 
although recent simulations generally find that central slopes are shallower 
than 1.5. 

This profile should characterise both the main halo
within which the Milky Way resides, and the subhalo cores that
survive within it. The central cusp of the main halo may be affected 
by subsequent stages in galaxy formation, however, since baryons
dominate the gravitational potential within the luminous part
of bright galaxies. In particular, processes like starbursts, galaxy mergers
or bar formation could heat and disrupt the dark matter cusp in
the centre of the Milky Way, M31 or the brighter members of the Local Group. 
Smaller subhaloes should be less strongly affected by baryonic 
processes, and thus they may preserve `pristine' dark matter cores, 
where the annihilation rate is still relatively high. Thus we will focus 
on comparing EGRET sources with small subhaloes, and ignore $\gamma$-ray
emission (or lack thereof) from the brightest members of the Local Group.

Using the semi-analytic model described in \cite{TaylorEtAl}, we have 
generated a number of realisations of halo substructure for a galaxy like 
the Milky Way. For each subhalo surviving at the present day, we calculated
the `flux multiplier' $J(\Psi)$ and the relative flux, as measured by an 
observer 8.5 kpc from the centre of the galaxy. A cut on subhalo masses 
of $>5\cdot10^{7} M_{\odot}$ was applied, as the models are incomplete
below this mass limit.

\begin{figure}
\vspace{7cm}
\includegraphics{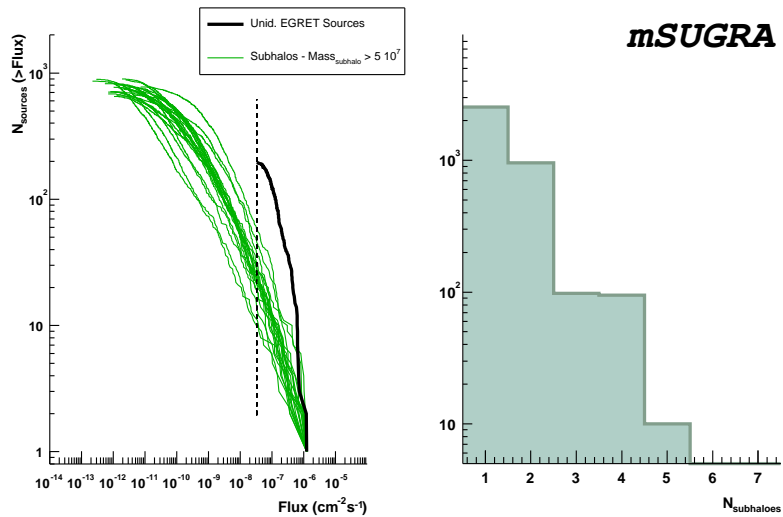}
\caption[]{(left) Constrains to the maximum number of subhaloes compatible to
unidentified EGRET sources. (right) Number distribution of all subhaloes with
fluxes detectable by EGRET in a mSUGRA scenario.}
\label{SubhalosFig}
\end{figure}

In Figure~\ref{SubhalosFig} (left) we compare the predicted fluxes from 
the subhaloes in the each of our model systems with the fluxes measured 
for the unidentified EGRET sources. For each realisation, we have scaled 
the luminosity of the brightest subhalo to the luminosity of the brightest 
EGRET source. This gives an upper limit on the emissivity of the dark matter,
for that particular model halo. By comparing the cumulative luminosity
functions, we can also determine how many of the unidentified sources could 
be subhaloes, assuming this scaling and a luminosity function like that of 
the model halo. This upper limit is between 10 and 58 of the EGRET sources, 
depending on the realisation, with a mean value of 26 $\pm$ 11. If we consider 
a neutralino of with $m_{\chi}=100$ GeV, we can compute the number of
gammas above the EGRET energy threshold with the \emph{Darksusy}
package \cite{GondoloetAl}. The $\left<\sigma v\right>_{\rm ann}$ needed to achieve
such fluxes are from $1\cdot10^{-25}$ to $5\cdot10^{-24}$
$cm^{3}s^{-1}$. These values exceed by one or two orders of magnitude
the allowed WMAP $\left<\sigma v\right>_{\rm ann}$ in some SUSY
scenarios. These numbers constrain an upper bound on the $\left<\sigma
v\right>_{\rm ann}$ for a 100 GeV neutralino.

As an example for a particular SUSY model, we consider to the set of 
mSUGRA scenarios discussed in \cite{PradaEtAl}. The range of models 
satisfying the WMAP relic density limits correspond to $\chi$ 
masses between approximately 70 GeV and 1400 GeV, while
$\left<\sigma v\right>_{\rm ann}$ lies in the range 1$\cdot$10$^{-29}$ 
to 3$\cdot$10$^{-26}$cm$^{3}$s$^{-1}$. For these SUSY models, we compute 
the expected $\gamma$-ray flux above 100 GeV for all 
simulated subhaloes. The Figure~\ref{SubhalosFig}(right) shows the number
distribution of all subhaloes with fluxes detectable by EGRET in this
mSUGRA scenario. The number of detectable subhaloes vary from 1 to
5. Most favourable models correspond to those neutralinos with
$\left<\sigma v\right>_{\rm ann}$ ranging between 3$\cdot$10$^{-26}$ cm$^{3}$s$^{-1}$
and 10$^{-27}$ cm$^{3}$s$^{-1}$ with masses from 100 to 400 GeV.

\section{Comparison with a Different Population of Halo Objects}

Given the small number of possible matches to between subhaloes and EGRET 
sources, it would be very useful to find another observable tracer of halo 
substructure. The High-Velocity Clouds (HVCs) are isolated concentrations 
of neutral hydrogen whose velocities distinguish them from the gas in 
the galactic disk \cite{BlitzEtAl}. The kinematics of these clouds 
suggest that they may orbit the Milky Way at large distances, or that 
they may even be distributed throughout the Local Group. While the Compact 
HVCs (CHVCs) seem the most promising tracers of dark substructure \cite{BraunBurton},
even the less compact clouds show similarities to subhaloes in their 
distribution, once cuts in halo distance are applied \cite{PutmanMoore}. 
The detection of a tidal interaction between a non-compact HVC and the 
dwarf galaxy LGS 3 further supports this association \cite{Robishaw}. 
Thus we will compare the distribution of both types of clouds with that of 
our model subhaloes, and with the distribution of EGRET sources.

\subsection{An All-sky Catalogue of High Velocity Clouds}

Two different catalogues of High Velocity Clouds are available for the
two hemispheres, and they can be combined to form an all-sky catalogue 
of HVCs. The northern LDS survey \cite{HeijBraun2} covered the sky north
of declination -30$^{0}$, while southern HIPASS survey
\cite{PutmanEtAl} covered the sky south declination +2$^{0}$. Both
surveys are different in terms of sensitivity, angular resolution,
velocity coverage and velocity resolution, but the same selection 
criteria was applied to both to identify isolated hydrogen clouds 
\cite{HeijBraun}. In Table~\ref{HIprop} we list the main parameters 
of each catalogue. Approximate positional accuracies were estimated by 
comparing published values for known dwarf galaxies with the positions 
of the associated hydrogen clouds.

\begin{table}[h] % 
\begin{tabular}{ccc}
\hline & LDS Survey & HIPASS Survey \\ \hline $V_{\rm LSR}$ coverage &
-450 $km s^{-1}$ to +400 $km s^{-1}$ & -700 $km s^{-1}$ to +500 $km
s^{-1}$ \\ V resolution & 1.0 $km s^{-1}$ & 26.4 $km s^{-1}$ \\
5$\sigma$ T Sensitivity & 70 mK & 1 mK \\ Angular resolution & 36' &
15.5' \\ Positional accuracy & 15' & 10' \\ \hline
\end{tabular}
\label{HIprop}
\caption[]{Observational properties for HIPASS and LDS HVCs surveys.}
\end{table}

\begin{figure}
\vspace{7cm}
\includegraphics{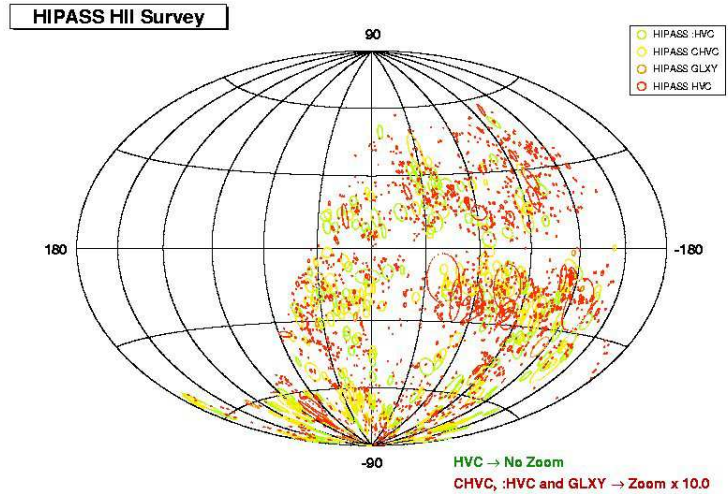}
\caption[]{Hydrogen clouds included in HIPASS survey.}
\label{HIIplots1}
\end{figure}

\begin{figure}
\vspace{7cm}
\includegraphics{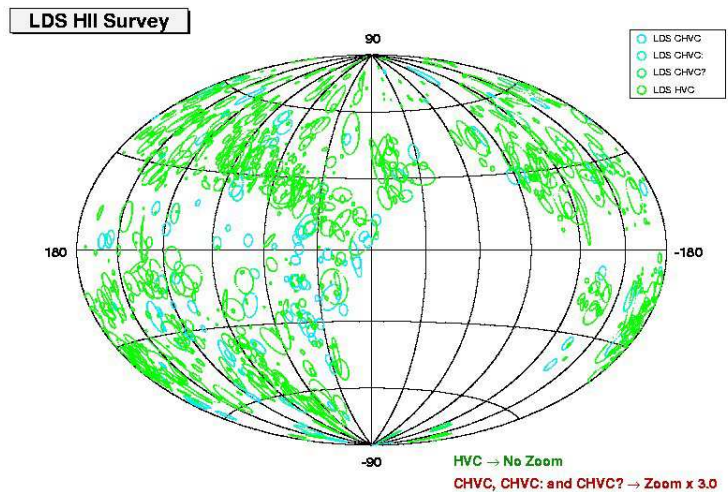}
\caption[]{Hydrogen clouds included in LDS survey.}
\label{HIIplots2}
\end{figure}

A total of 917 clouds were catalogued in the northern LDS survey, of
which 777 are HVCs, 90 are CHVCs (isolated clouds in position and
velocity), 37 were designated as CHVC: and 13 designated as CHVC?
(clouds which were not unambiguously designated as CHVCs by the criteria
described in \cite{PutmanEtAl}). For the HIPASS southern survey, a total of
1997 clouds were detected, of which 41 are actually galaxies (GLXY), 1618 are
HVCs, 179 are CHVCs and 159 are designated as :HVC as they could not be
unambiguously classified as CHVCs because their elongation towards low level
extended emissions. All the hydrogen clouds detected by either survey are
shown in Figure~\ref{HIIplots1} and Figure~\ref{HIIplots2}.

\subsection{Positional Coincidences Between HVCs and Unidentified EGRET Sources}

Given the HVC positions from the HIPASS and LDS catalogues, we can 
test for positional coincidence between hydrogen clouds and $\gamma$-ray
sources. More specifically, we can compare the positions of the maximum 
column density in neutral hydrogen for each cloud with the positions of 
the unidentified EGRET sources, on the supposition that both trace the
dynamical centres of bound dark matter subhaloes. In order to determine
the significance of positional coincidences, we define a relative distance
between each hydrogen cloud and the each of the unidentified EGRET sources as:

\begin{eqnarray}
\delta^{2} = \frac{|\vec{x}_{\rm egret} - \vec{x}_{\rm
cloud}|^{2}}{\sigma^{2}_{\rm egret} + \sigma^{2}_{\rm cloud}}
\end{eqnarray}

where $\vec{x}_{\rm egret}$ and $\vec{x}_{\rm cloud}$ are the galactic
longitude and latitude of the unidentified EGRET sources and the
hydrogen clouds, expressed in degrees. We take as an characteristic 
uncertainty in the EGRET positions $\sigma_{\rm egret}=\theta_{95}/2$, 
where $\theta_{95}$ is the 95$\%$ EGRET detection confidence contour.
For $\sigma_{\rm cloud}$ we choose the positional accuracies of HIPASS 
and LDS surveys given above. As $\sigma_{\rm egret} > \sigma_{\rm cloud}$, 
$\delta$ values will be dominated by $\sigma_{\rm egret}$.

\begin{figure}[h]
\vspace{7cm}
\includegraphics{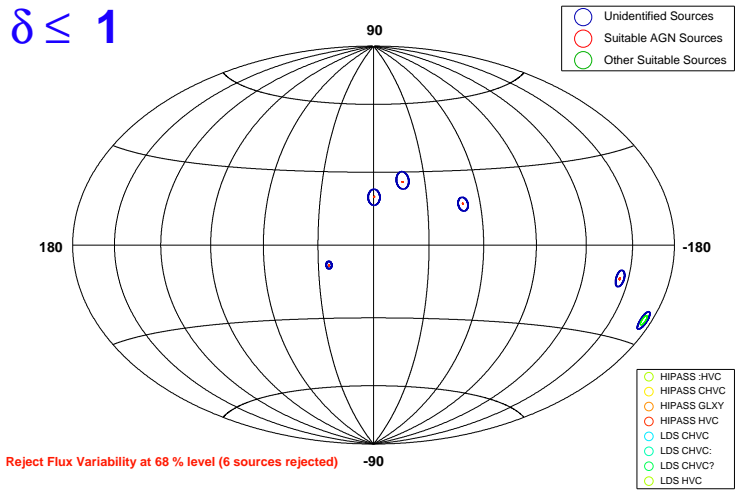}
\caption[]{Unidentified EGRET sources whose positions coincide 
 with HVCs at $\delta\leq1$, after applying 68$\%$ flux variability
 rejection.}
\label{TheSources}
\end{figure}

A total of 10 HIPASS HVCs and 2 LDS HVCs have an unidentified EGRET
source lying within $\delta\leq$1 of their point of maximum emission. 
As the number of unidentified EGRET sources in the field of view of 
HIPASS(LDS) surveys are quite similar, the different number of coincidences 
found for the two catalogues may be explained because the better sensitivity 
of HIPASS survey (HIPASS contains roughly double HI clouds than LDS) and 
worst LDS positional accuracies.

The $\gamma$-ray emission from neutralino annihilation should be steady on 
observationally accessible timescales. The variability of all sources 
in the 3rd EGRET Catalogue has been evaluated, assuming a model of the 
flux distribution to characterise the flux variation, and using both 
positive flux detections and upper limits \cite{Nolan}. This method
is quite successful at distinguishing between  between the known (variable)
quasars and the known (non-variable) pulsars.
As we are only interested in those steady $\gamma$-ray emitters,
independent of how much variable they are, we make use of the
$V_{12}$ parameter, which indicates the confidence with which 
the hypothesis of variability can be rejected for a given source 
(100$\%$ = steady source). If we apply a cut at the 68$\%$ level to
our $\delta\leq$1 sample, 6 sources are rejected, resulting in
5 (1) remaining coincidences between HVCs to and unidentified EGRET 
sources for the HIPASS (LDS) surveys. These sources are shown in 
Figure~\ref{TheSources} and listed in Table~\ref{HIItoEGRET}.

To check for significance of these coincidences, we perform several
simulations in which we distribute $10^{4}$ fake EGRET sources isotropically 
and evaluate the probability of finding coincident HVCs and EGRET sources, 
after including the rejection factor from variability. The average number 
of random associations found were $1.0\pm0.3$ for LDS survey and 
$3.1\pm0.5$ for HIPASS survey, for the same coincidence cut criteria
of $\delta\leq$1. Thus the real result obtained for HIPASS clouds is 
only marginally (4$\sigma$) above the level expected from random 
coincidence.

\begin{table}[h] 
\begin{tabular}{cccccc}
\hline
\small {\bf 3rd EGRET Name} & \small {\bf HII Name} & \small {\bf $\sigma$}\\

\hline

\small  J0439+1105 & \small HVC 186.2-22.8-206 $\ddagger$ & {\bf 0.18}\\
\small  J0616-0720 & \small HVC 215.8-11.2+162 $\dagger$ & {\bf 0.54}\\
\small  J1329-4602 & \small HVC 310.0+16.5+182 $\dagger$ & {\bf 0.51}\\
\small  J1527-2358 & \small HVC 343.1+25.9-144 $\dagger$ & {\bf 0.98}\\
\small  J1635-1751 & \small HVC 359.6+19.7-149 $\dagger$ & {\bf 0.32}\\
\small  J1904-1124 & \small HVC 024.1-08.2-238 $\dagger$ & {\bf 0.48}\\

\hline
\end{tabular}
\label{HIItoEGRET}
\caption[]{Unidentified EGRET Sources and HVCs Found to be Spatially
Coincident. $\dagger$ = HIPASS Survey, $\ddagger$ = LDS Survey.}
\end{table}
\normalsize

\section{Conclusions}

If the dark matter particle is a supersymmetric neutralino, one of the most
popular candidates currently under consideration, then local concentrations 
of dark matter may be detectable as $\gamma$-ray sources distributed throughout
the halo of the Milky Way. In this paper, we have compared the luminosities of 
unidentified sources in the EGRET $\gamma$-ray catalogue with the luminosities
of local dark matter substructure predicted by a semi-analytic model of 
neutralino annihilation in the halo of the Milky Way. We have also looked
for spatial coincidences between EGRET sources and high-velocity neutral
gas, which could be an observational tracer of local dark matter.

We find that most of the unidentified EGRET sources cannot be dark matter
subhaloes. Whatever the emissivity of annihilating dark matter, the 
cumulative luminosity function for the two populations is sufficiently 
different that only $\sim$26 $\pm$ 11 of the EGRET sources could correspond 
to local substructure. Applying further cuts to sources on the basis of 
location or variability should place an even stronger constraint on the 
number of possible sources, and thus on the properties of dark matter.

We also find that most of the HVCs cannot correspond to EGRET sources 
either. While there are a few close positional coincidences between the 
two populations, the overall number is roughly that expected from random 
superposition. Thus there is no indication that the EGRET sources are 
associated with HVCs, whether or not the latter trace dark matter. This 
is interesting, because HVCs could also produce $\gamma$-rays through other, 
less exotic mechanisms, such as interactions with cosmic rays. Comparing 
HVCs with sources in lower energy catalogues might produce a stronger 
constraint on this mechanism, however, since the spectrum of $\gamma$-rays 
produced in this case is broader than that expected for $\chi$ annihilation.

\end{article}
\end{document}